\documentclass[aip,jap,reprint]{revtex4-1}
\draft 
\usepackage{amsmath}
\usepackage{amssymb}
\usepackage{graphicx}
\usepackage{tikz}
\usepackage{multirow}
\begin{document}
\title{Improved Nb SIS devices for heterodyne mixers between 700~GHz and 1.3~THz 
with NbTiN transmission lines using a normal metal energy relaxation layer}
\author{M.~P.~Westig}
\email[]{westig@ph1.uni-koeln.de}
\affiliation{K\"olner Observatorium f\"ur Submillimeter Astronomie 
(KOSMA), I.~Physikalisches Institut, 
Universit\"at zu K\"oln, Z\"ulpicher Stra\ss e 77,
D-50937 K\"oln, Germany}
\author{S.~Selig}
\affiliation{K\"olner Observatorium f\"ur Submillimeter Astronomie 
(KOSMA), I.~Physikalisches Institut, 
Universit\"at zu K\"oln, Z\"ulpicher Stra\ss e 77,
D-50937 K\"oln, Germany}
\author{K.~Jacobs}
\affiliation{K\"olner Observatorium f\"ur Submillimeter Astronomie 
(KOSMA), I.~Physikalisches Institut, 
Universit\"at zu K\"oln, Z\"ulpicher Stra\ss e 77,
D-50937 K\"oln, Germany}
\author{T.~M.~Klapwijk}
\affiliation{Kavli Institute of NanoScience, Faculty of Applied Sciences, Delft University of 
Technology, Lorentzweg 1, 2628 CJ Delft, The Netherlands}
\author{C.~E.~Honingh}
\affiliation{K\"olner Observatorium f\"ur Submillimeter Astronomie 
(KOSMA), I.~Physikalisches Institut, 
Universit\"at zu K\"oln, Z\"ulpicher Stra\ss e 77,
D-50937 K\"oln, Germany}
\date{\today}
\begin{abstract}
In this paper we demonstrate experimentally the implementation of a niobium-trilayer 
junction with an aluminum-oxide tunnel barrier, embedded in a high-gap 
superconducting niobium-titanium-nitride circuit. Previously reported heating by 
quasiparticle trapping is removed by inserting a normal metal layer 
of gold between the niobium junction and the niobium-titanium-nitride 
layer. We analyze in \emph{dc}-characterization measurements the cooling of the nonequilibrium quasiparticles 
in various device geometries having different gold layer thickness and shape. 
Our work is concluded with remarks for future heterodyne mixer experiments using our device
technology.
\end{abstract}
\pacs{85.25.-j, 74.78.-w, 85.25.Pb}
\maketitle 
\section{Introduction}
\label{sec:01}
Low-noise heterodyne receivers working around 1~THz are highly desired instruments
in submillimeter (submm) radio astronomy. They furnish a unique
spectral resolution of the order of $\nu/\delta\nu \approx 10^{5}-10^{6}$ 
which allows astronomers to resolve complicated emission spectra of 
various astronomical sources. 
Detection of molecular or atomic THz emission lines provides 
access to a rich information source containing the physical and chemical 
conditions of hot molecular/atomic gas which plays a role in star formation in the 
interstellar medium. Precise measurements of these conditions allow, 
therefore, to complete the fundamental idea how stars form and evolve.\citep{stahler2005} 
For this purpose it is necessary to have the most 
sensitive detectors available which are the centerpiece of every heterodyne receiver system.

High sensitivity heterodyne receivers for radio-astronomy employ two different types of 
superconducting detectors. For the submm frequency range of 0.3-1.3~THz, 
superconductor-insulator-superconductor (SIS) frequency mixers are used. Here, one uses in 
practice the high-quality niobium-trilayer technology in combination with an 
aluminum-oxide ($\mathrm{AlO_{x}}$) or aluminum-nitride (AlN)\citep{zijlstra2007} 
tunnel barrier, resulting in 
junctions with the lowest subgap currents. In SIS heterodyne mixers the highest detectable 
frequency is set by the superconducting gap energy $\Delta$ of the junction's electrode 
material. For a symmetric junction this frequency is $\nu_{max} = 4\Delta/h$, where 
$2\Delta$ is the superconducting pair breaking energy of one electrode and $h$ is 
the Planck constant. Heterodyne mixing well above the pair breaking frequency 
of the single electrodes is not intrinsically limited by the electrode material itself, 
albeit that in the frequency range $2\Delta/h - 4\Delta/h$ excess quasiparticles are 
created, absorbing a part of the incoming signal. For the higher frequencies 
superconducting hot-electron bolometer mixers (HEB) are used.\citep{zmuidzinas2004} 
In an HEB device there is in principle no limitation with respect to the detection 
frequency, however, it has a lower intermediate frequency bandwidth than a 
SIS mixer. SIS and HEB devices have provided a remarkable success of 
ground-based,\citep{guesten2006,kooi2009,wootten2009} 
airborne\citep{heyminck2012,puetz2012} and space\citep{degraauw2010_short} 
observatories and will be of significant importance for future astronomical observations.

Since heterodyne detection is a coherent detection technique where amplitude
and phase of the incoming signal are preserved during the frequency mixing process, 
the Heisenberg uncertainty principle imposes a fundamental noise limit on the 
device sensitivity.\citep{caves1982} This noise limit corresponds to the photon 
shot noise at the input of the mixer coming from a background of one photon 
per second per Hertz of bandwidth. When expressed as a noise 
temperature, this results in $T_{qn} = h\nu/k_{B}$, i.e.~0.047~K/GHz, also 
referred to as the quantum noise limit, where $\nu$ is the signal frequency 
and $k_{B}$ is the Boltzmann constant. 

Quantum limited mixer performance\citep{tucker1985} is reliably achieved up to frequencies of about 
680~GHz, the superconducting gap frequency of the niobium (Nb) embedding circuit. 
For larger frequencies, substantial ohmic losses are observed 
in the embedding circuit due to breaking Cooper pairs in the material in 
combination with the usually large normal state resistance of superconductors.
Possible solutions to minimize these losses are embedding 
circuits consisting of different materials than Nb.

\emph{Jackson et al.}\citep{jackson2001,jackson2005} 
reported receiver noise temperatures
of approximately 200~K at 850~GHz and 400~K at 
1~THz for a Nb SIS mixer using a hybrid embedding circuit consisting of a niobium-titanium-nitride 
(NbTiN) bottom layer and an Al wiring layer. \emph{Bin et al.}\citep{bin1996} demonstrated a Nb SIS 
mixer using a double normal-conducting embedding structure made of Al, however, they obtained a higher 
receiver noise temperature of approximately 840~K at 1042~GHz due to the \emph{rf}-signal 
loss in the Al. Recently, \emph{Wang et al.}\citep{wang2013} fabricated high-quality, 
high-current density niobium nitride (NbN) SIS junctions with aluminum-nitride tunnel barriers 
where the complete embedding circuit also consists of the high-gap superconductor NbN.

In this previous reported work it was shown to be important to avoid heating effects of the 
electron gas in the device. In this article we propose a different solution for this particular 
problem and show its effectiveness. Device heating by quasiparticle trapping is removed 
by inserting a normal metal layer of gold (Au) between the Nb junction and the NbTiN 
embedding circuit where nonequilibrium quasiparticles relax their energy.

In the following section we present the conceptional background of our device design and a detailed explanation
of the expected quasiparticle processes in our device. Section~\ref{sec:03} describes the fabrication and
the measurement setup. In Section~\ref{sec:04} we show results of extensive 
\emph{dc}-characterization measurements 
of various device geometries and present our analysis. Section~\ref{sec:05} discusses possible adverse
effects of the Au layer in the superconducting device before Sec.~\ref{sec:06} concludes the paper.
\section{Design considerations and conceptual background}
\label{sec:02}
In SIS junctions for heterodyne mixer applications one demands a nearly ideal nonlinearity of the 
I,V characteristic: a low subgap current and a sharp onset of the quasiparticle current branch. For a 
low subgap current one needs a high-quality tunnel barrier, which in practice is best achieved with 
Nb-trilayer technology with an $\mathrm{AlO_{x}}$ or AlN tunnel barrier. The frequency range from 
700 GHz to 1.3 THz can be covered by such a nonlinear I,V curve provided one can avoid or minimize 
absorption of the signal by the embedding circuit. If this is the case one ought to be able to obtain
high-performance devices reaching quantum limited noise temperatures.\citep{tucker1985}
The preferred solution is the use of a superconductor with 
a higher critical temperature and a higher energy gap than Nb. However, it has 
been found that the power generated by the tunnel current/voltage cannot effectively be removed and 
nonequilibrium quasiparticles begin to play a role. 

For a SIS junction with identical electrode materials, 
i.e.~a symmetric junction, $I_{qp}$ is given by\citep{tinkham2004}
\begin{equation}
\label{eq:01}
\begin{split}
I_{qp}(V) = \frac{1}{eR_{N}} \int_{-\infty}^{+\infty}dE~&N_{S}(E,\Delta)N_{S}(E+eV,\Delta)\\
&\cdot\left[f(E,T) - f(E+eV,T)\right]~.
\end{split}
\end{equation}
Here $V$ is the bias voltage applied to the two electrodes of the SIS junction and 
$f(E,T)$ is the Fermi distribution function of quasiparticles with energy $E$, 
having a temperature $T = T_{e}$. The subscript "$e$`` emphasizes that this is the electron temperature. 
The quantities $N_S$ are the normalized BCS density of states: 0 for $|E|< \Delta$ and 
$\lvert E \rvert/(E^2-\Delta^2)^{1/2}$ for $\lvert E \rvert > \Delta$.\citep{bardeen1957} Here, $\lvert\Delta\rvert$
is the energy gap in the quasiparticle excitation spectrum for electronlike and holelike quasiparticles. An 
energy of at least $2\Delta$ is needed in order to destroy a Cooper pair and
create two quasiparticle excitations. In the Nb-trilayer technology one of the 
electrodes is actually Nb with a thin layer of aluminum, i.e.~represented by a 
proximity-induced density of states which is close to the Nb density of states, 
although slightly modified. 

In a realistic I,V curve the singularity in the density of states leads to a sharp rise at the gap voltage $V_{g}=2\Delta/e$. 
However, the sharpness is experimentally often less than is consistent with a real singularity. 
In practice, this can have a number of reasons and is usually taken into account by a phenomenological 
broadening parameter of the singularity, $\Gamma$. The superconducting density of states $N_S$ in 
Eq.~(\ref{eq:01}) is then expressed as\citep{dynes1978} 
\begin{equation}
\label{eq:02}
N_{S}(E,\Delta) = \mathrm{Re}\left(\frac{E-i\Gamma}{\sqrt{(E - i\Gamma)^{2} -\Delta(T)^{2}}}\right)~.
\end{equation}
This expression suggests that one assumes that the actual density of states of the superconducting 
electrodes is modified. To avoid this misunderstanding we call $\Gamma$ a \emph{phenomenological} 
parameter, which serves to deal with non-idealities in the sharpness of the I,V curve 
compared to the idealized situation described by Eq.~(\ref{eq:01}). 
The rich variety of all other possible origins of non-idealities not contained in Eq.~(\ref{eq:01}) 
will be further ignored.

Since one often prefers in practice tunnel barriers which have a high transmissivity, i.e.~low 
$R_{N}A$-values or high critical-current densities, we emphasize that Eq.~(\ref{eq:01}) represents the 
tunneling current assuming that only lowest order tunnel processes are relevant.  As long as the 
tunnel probability is much less than unity, the higher order processes are indeed negligible. However, 
for very thin tunnel barriers, needed for broadband heterodyne mixers, 
the likelihood of 'weak spots' increases causing a fraction of the tunnel barrier to have a higher tunneling 
probability, even close to unity, allowing significant contributions from higher order tunneling processes. 
These are easily identified as excess shot noise,\citep{dieleman1997} and are also visible as excess 
subgap currents, neglected by Eq.~(\ref{eq:01}). 

In principle, in using tunnel junctions it is assumed that the tunnel barrier is a weak disturbance 
i.e.~that the two electrodes are undisturbed by the tunneling current. This is reflected in Eq.~(\ref{eq:01}) by the 
use of the Fermi-Dirac distribution at the temperature $T$ as well as the energy gap $\Delta$ at the same 
temperature. 

In this article we are interested in SIS devices in which this particular assumption is no longer 
satisfied. We will focus on the option that due to the dissipated power, the temperature $T_{e}$ of the electron system is 
higher than the bath temperature. This new temperature $T_{e}$ will enter the Fermi-functions in Eq.~(\ref{eq:01}). 
However, it will also influence the energy gap $\Delta$, which is itself also controlled by the Fermi-function through
the self-consistency relation\citep{bardeen1957}
\begin{equation}
\label{eq:03}
\frac{1}{N(E_{F})\mathcal{V}} = \int_{0}^{k_{B}\theta_{D}}d\epsilon~
\frac{1-2f\left[(\epsilon^{2} + \Delta(T_{e})^2)^{1/2}\right]}{(\epsilon^{2} + \Delta(T_{e})^2)^{1/2}}~,
\end{equation}
where $N(E_{F})$ is the single-spin density of states at the Fermi 
energy in the normal state, $k_{B}$ is the Boltzmann constant, 
$\theta_{D}$ is the Debye temperature of the material, $\mathcal{V}$ is the average attractive 
interaction potential of superconductivity describing phonon exchange between 
electrons\citep{bardeen1957} and $\epsilon$ is the independent 
quasiparticle energy measured relative to the Fermi energy. Hence, Eq.~(\ref{eq:03}) is an implicit expression 
relating $\Delta$ to $T_{e}$. The temperature $T_{e}$ is dependent on the power generated by the 
tunneling process, which rapidly rises for bias voltages near the gap voltage $V_g$, implying 
that the relevant energy gap for the tunneling decreases leading to the possibility of back-bending 
(a negative slope of $I_{qp}$ around $V_g$, compare with figure~\ref{fig:04}(a)).  

The actual temperature $T_e$ in comparison to the bath temperature, the phonon-temperature 
$T_{ph}$, depends on two processes.

First, a thin tunnel barrier leads to a higher density of hot electrons injected per unit time. We assume 
that the electron-electron interaction time  $\tau_{e-e}$ is short enough to establish that an effective 
energy relaxation will occur by electron-phonon interaction, assuming  $\tau_{e-e} < \tau_{e-ph}$, if 
one has a closed volume. Otherwise, the outdiffusion of hot electrons compensated by the indiffusion 
of cool electrons establishes equilibrium. For conventional aluminum oxide barriers the barrier integrity 
breaks down before nonequilibrium processes become relevant. However, for lower  
$R_{N}A$-products, such as is possible by using AlN barriers, an elevated electron temperature does play 
a role. In addition, it might play a role with more disordered superconducting materials, such as NbN, 
where the diffusion coefficient is much smaller than in Nb and Al.   

\begin{figure}[t!]
\centering
\includegraphics[width=1\columnwidth]{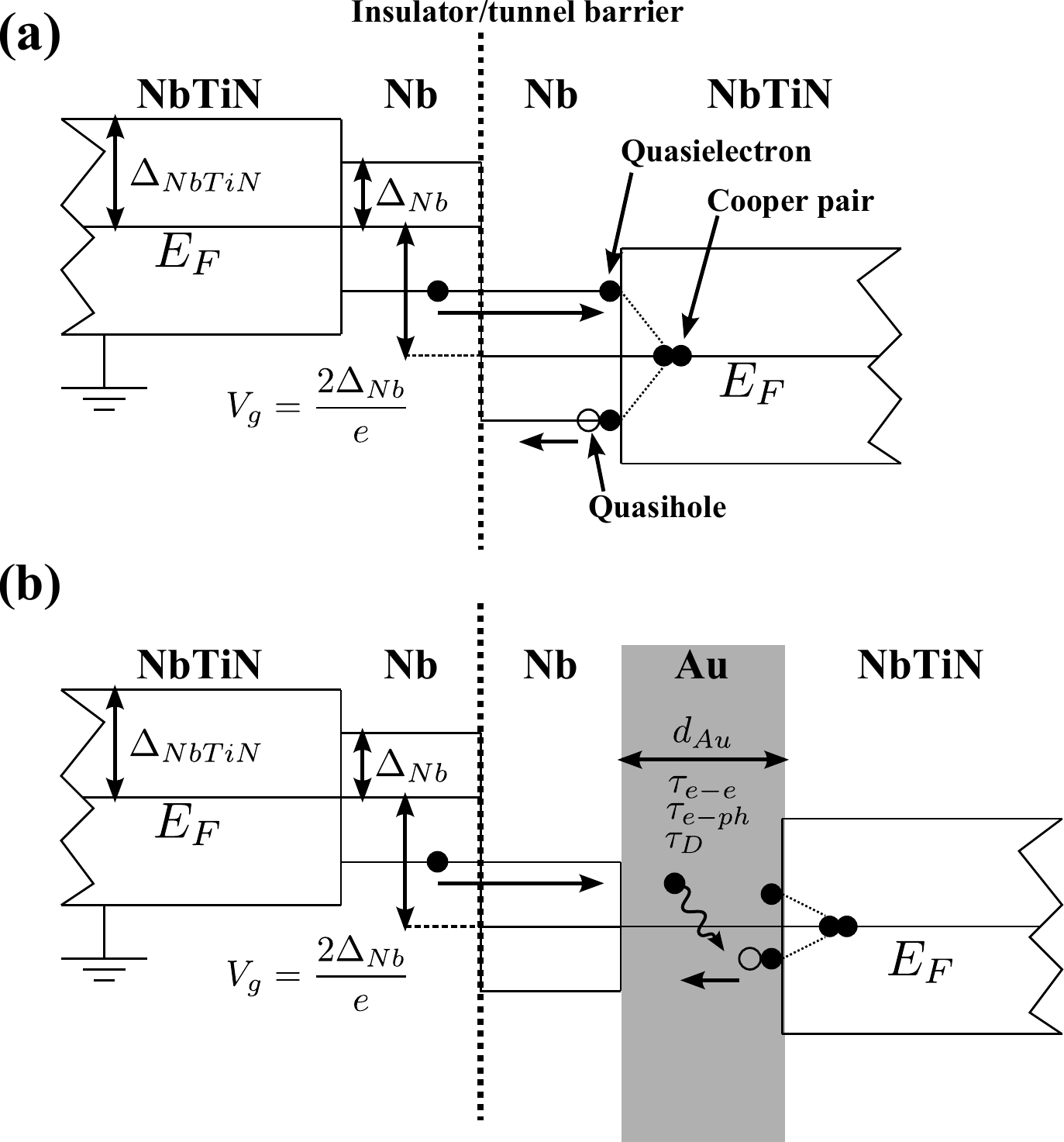}
\caption{(a) Energy diagram of the control device (Fig.~\ref{fig:02}(a), \#1) with applied bias
voltage $V_{g}$, inspired by \emph{Leone et al.}\citep{leone2000}
(b) shows the modification of the energy
diagram when a normal metal Au layer (shaded area) is inserted between the Nb and NbTiN layer, having
no superconducting gap in the electronic excitation spectrum. 
Here, a quasiparticle injected from the tunnel barrier into the Au layer thermalizes with other quasiparticles
after a time $\tau_{e-e}$ and becomes an excited quasiparticle with energy $\epsilon = \Delta_{Nb}$ 
which is relaxed by inelastic electron-phonon scattering after the time $\tau_{e-ph}$. 
$d_{Au}$ indicates the thickness of the layer and $\tau_{D}$ is the diffusion time through the structure.}
\label{fig:01}
\end{figure}
Secondly, in Nb SIS junctions, embedded between higher-gap superconductors such as NbTiN, 
outdiffusion is blocked by the \emph{Andreev} trap formed between the tunnel barrier and the high 
energy gap of the NbTiN. This leads to a geometrical trapping of the 
quasiparticles,\citep{leone2000} illustrated in Fig.~\ref{fig:01}(a). Therefore, although charge is 
transported from the Nb SIS junction into the NbTiN, the energy transport is blocked, a process 
known as \emph{Andreev} reflection.\citep{andreev1964} In this case, even for high $R_NA$-value SIS 
devices, the increased electron temperature plays a major role.\citep{leone2000}

The best possible solution for high quality SIS mixing is, nevertheless, the use of high-gap 
superconducting transmission lines in combination with Nb SIS devices. Therefore, we seek 
a solution to avoid these nonequilibrium processes in an attempt to maintain the electron 
temperature $T_{e}$ as close as possible to the phonon temperature $T_{ph}$. Our 
solution is to insert between the Nb film and the NbTiN film a normal metal Au
layer, which allows quick thermalization of the hot electrons in the Nb as shown in 
Fig.~\ref{fig:01}(b). Two possibilities have been investigated in this paper. Case I, 
in which the normal metal layer is laterally confined to the same dimensions as the 
Nb. And Case II, in which the normal metal layer is substantially wider than the 
Nb electrode offering a larger volume for a process which we 
call \emph{geometrically assisted cooling}.     

In Case I with the confined normal metal layer, energy relaxation occurs through 
electron-electron, electron-phonon and phonon-escape processes with 
characteristic time $\tau_{es}$. In order to quantitatively estimate the effect of 
the dissipated power $P$ on $T_{e}$, we assume a linearized heat balance 
equation \citep{skocpol1974} with a heat-transfer coefficient $\alpha$:
\begin{equation}
\label{eq:04}
P = \alpha \left(T_{e} - T_{ph}\right)~.
\end{equation}

In Case II a different theoretical description is used, which accounts for the 
outdiffusion of nonequilibrium quasiparticles from the Nb junction into the wide Au layer. 
For this purpose we use the following heat balance equation\citep{skocpol1974}
\begin{equation}
\label{eq:05}
-\kappa \left[r^2\frac{d^2T_{e}(r)}{dr^2} + r\frac{dT_{e}(r)}{dr}\right] + 
\frac{Y}{d_{Au}} r^2 \left[T_{e}(r)-T_{ph}\right] = 0~.	
\end{equation}

Its solution is the radial temperature distribution around the SIS junction in the 
Au layer. The thermal conductivity is $\kappa$, $r$ is the radial
position, measured from the center of the SIS junction, $T_{e}(r)$ is the electron 
temperature, $Y$ is the heat-transfer coefficient (measured in $\mathrm{W/m^2 K}$) 
to the phonon system, $T_{ph}$ is the bath (phonon) temperature as before and 
$d_{Au}$ is the thickness of the Au layer. When $Y$ describes the heat transfer 
to the metallic phonon system, then $Y = Y_{e-ph}$ with $Y_{e-ph}$ being the 
electron-phonon heat-transfer coefficient. On the other hand, when $Y$ 
describes heat transfer to the phonon bath system (substrate or liquid helium (LHe)), 
then $Y = Y_{K}$ with $Y_{K}$ being the Kapitza conductance between the metal and the substrate 
or between the metal and the LHe.
\section{\label{sec:03}Device variations and measurement setup}
\begin{table*}[t!]
\caption{\label{tab:01} Gold layer thickness $d_{Au}$, gold volume $V_{Au}$, critical current 
density $j_{c}$, normal state resistance $R_{N}$, junction area $A$ and $R_{N}A$-product of 
S'SISS' device \#1 and S'SISNS' devices \#2-\#7 (compare with Fig.~\ref{fig:02}). The critical current density of 
devices \#1, \#2, \#3 and \#6 is determined via the measured junction area, the normal state 
resistance and the gap voltage (compare with the gap energies in Table~\ref{tab:04}) by using the
\emph{Ambegaokar and Baratoff} theory.\citep{ambegaokar1963_01}
On the other hand, the small areas of devices \#4 and \#7 are more accurately determined 
under the assumption that they have the same current 
density like \#3 and \#6 since they were fabricated on the same wafer. Therefore, we mark their $j_{c}$
and $A$ values with an asterisk (*). For device \#5 
(also marked with an asterisk) the area is equally determined via a current density 
measurement of a larger area device, not indicated in this table. Devices \#1 - \#7 were 
directly immersed in LHe, whereas the $j_{c}$ values of the SNS' devices \#8 and \#9 were measured 
with the devices mounted on the 4.2~K stage of a LHe dewar. The SNS' device areas 
are determined with the assumption that the measured $R_{N}$ is most likely the Sharvin 
resistance.\citep{sharvin1965}}
\begin{ruledtabular}
\begin{tabular}{l l l l l l l l}
\#&Type&$d_{Au}$~[nm]&$V_{Au}~[\mu\mathrm{m}^3]$&$j_{c}$~[kA/cm$^{2}$]
&$R_{N}~[\Omega]$ &$A~[\mu\mathrm{m^2}]$&$R_{N}A~[\Omega \mu\mathrm{m^2}]$\\
\hline
1&S'SISS'&-&-&6.9&13.6&2.25&30.6\\
\hline
2&\multirow{6}{*}{S'SISNS'}&20&0.045&8.2&11.42&2.23&25.5\\
3&&80&0.2&11.9&6.95&2.52&17.5\\
4&&80&0.056&$(11.9)^{*}$&25.9&$(0.7)^{*}$&\\
5&&20+60&1.3&$(15.4)^{*}$&11.45&$(1.18)^{*}$&\\
6&&120&8.36&6.4&3.65&9.19&33.5\\
7&&120&4.86&$(6.4)^{*}$&30.59&$(1.08)^{*}$&\\
\hline
8&\multirow{2}{*}{SNS'}&80&&$2.6\cdot 10^{4}$&1.76&$7.9\cdot 10^{-4}$&\\
9&&120&&$2.1\cdot 10^{4}$&0.082&$2.5\cdot 10^{-2}$&
\end{tabular}
\end{ruledtabular}
\end{table*}
\subsection{\label{sec:03a}Fabrication}
\begin{figure}[t]
\centering
\includegraphics[width=\columnwidth]{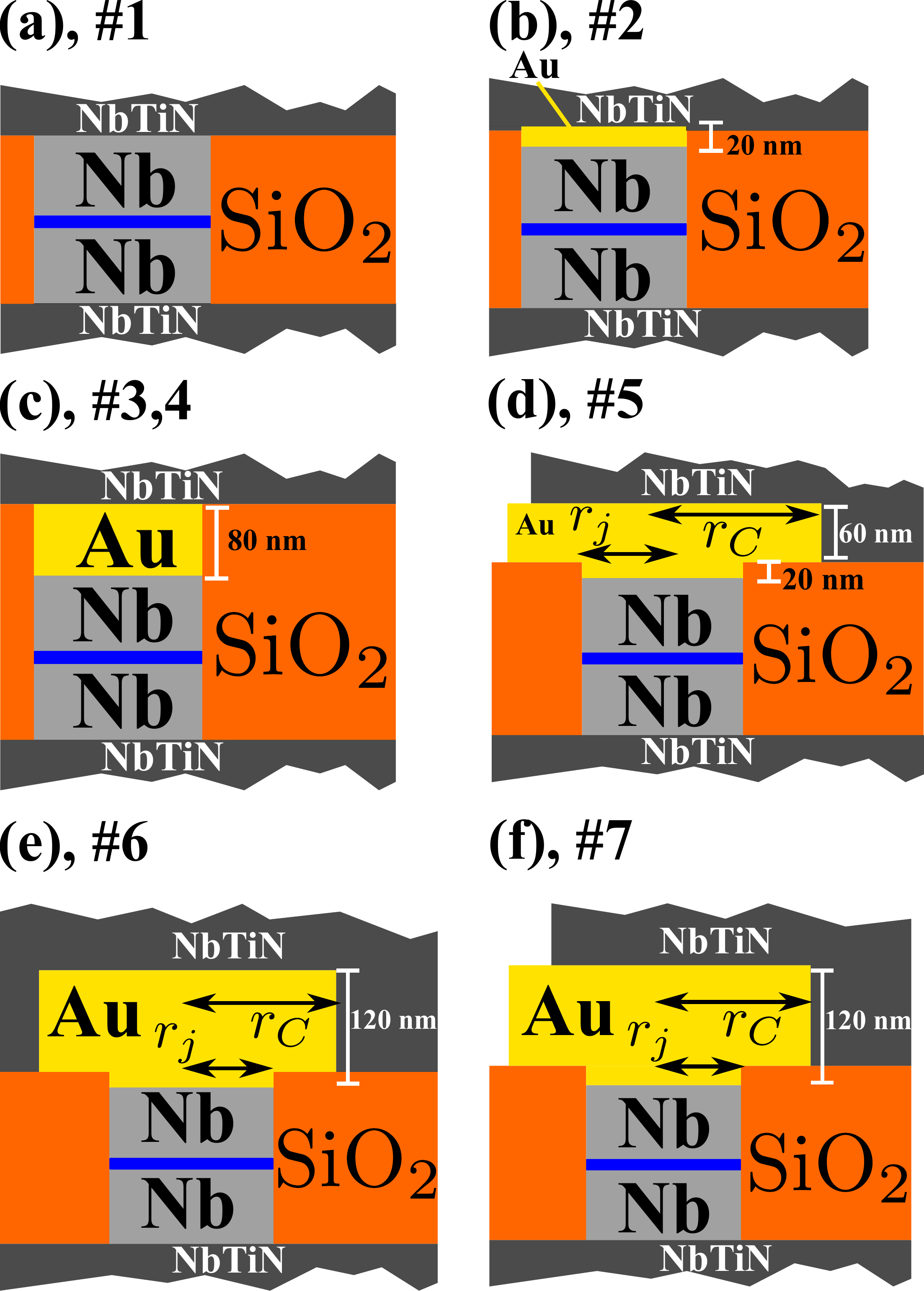}
\caption{(Color online) Figures (a)-(f) (devices \#1-\#7) show the device layouts 
analyzed in this paper, having Au layers of various thickness $d_{Au}$ and radius $r_{C}$ whereas 
the junction radius is indicated by $r_{j}$. The layer sequence for device \#1 is S'SISS' 
(S=Nb, S' = NbTiN and I = $\mathrm{AlO_{x}}$ (insulator, blue layer)) whereas for the other devices 
the sequence is S'SISNS' (N = Au). In the figures we omit to show the silicon handle wafer.
Devices (a)-(c) correspond to Case I whereas (d)-(f) correspond to Case II.}
\label{fig:02}
\end{figure}

The $\mathrm{S'SISS'}$ and $\mathrm{S'SISNS'}$ devices ($\mathrm{S'} = \mathrm{NbTiN}$, 
$\mathrm{S} = \mathrm{Nb}$, I = $\mathrm{AlO_{x}}$ and $\mathrm{N} = \mathrm{Au}$) 
are fabricated on silicon substrates having a thickness of $525~\mu$m (omitted in Figs.~\ref{fig:02}(a)-(f)).
The first wafer contains control devices which are fabricated 
without the Au layer (compare with device \#1 in Table~\ref{tab:01} and Fig.~\ref{fig:02}(a)) and serve as reference in 
the data interpretation. The $\mathrm{NbTiN/Nb/AlO_{x}/Nb}$ layers
are deposited by \emph{dc}-magnetron sputtering and are patterned by UV optical lithography.
The bottom layer is room temperature sputtered NbTiN of thickness 350~nm and 
the Nb electrodes have a thickness of 100~nm each. The tunnel junction areas are defined by electron 
beam lithography (EBL) and reactive ion etching (RIE) of the $\mathrm{NbTiN/Nb/AlO_{x}/Nb}$ layers.
In a next step, an \emph{rf}-sputtered $\mathrm{SiO_{2}}$ dielectric of 200-300~nm thickness is 
patterned by self-aligned liftoff. Finally, a 400~nm thick room 
temperature sputtered NbTiN wiring layer is defined by UV optical lithography.

For the other devices, the same process steps as described before are used
except for specific details which are summarized below.

A second and a third wafer include an additional Au layer of 20 and 80~nm thickness. 
The $\mathrm{NbTiN/Nb/AlO_{x}/Nb/Au}$ layers are
deposited by \emph{dc}-magnetron-sputtering and are patterned by UV optical lithography
(compare with devices \#2-\#4 in Table~\ref{tab:01} and Figs.~\ref{fig:02}(b) and (c)). 
The tunnel junction areas are again 
defined by EBL and RIE of the $\mathrm{NbTiN/Nb/AlO_{x}/Nb/Au}$ layers.
For the wafer containing the 80~nm thick Au layer, chemical mechanical polishing (CMP)
is used to planarize the $\mathrm{SiO_{2}}$ dielectric in order to enhance the quality of the
subsequently performed lift-off.

A fourth wafer contains additional devices with a total Au layer thickness of 80~nm. 
In contrast to the wafer before, on this wafer the shape of the Au layer was varied. First a
$\mathrm{NbTiN/Nb/AlO_{x}/Nb/Au}$ layer with an Au layer thickness of only 20~nm 
is deposited by \emph{dc}-magnetron sputtering and is again patterned by 
UV optical lithography (compare with device \#5 in Table~\ref{tab:01} and Fig.~\ref{fig:02}(d)). 
After the definition of the tunnel junction areas by EBL and RIE of the 
$\mathrm{NbTiN/Nb/AlO_{x}/Nb/Au}$ layers and deposition
of the $\mathrm{SiO_{2}}$ dielectric layer, an additional Au layer (cap) having a 
thickness of 60~nm and a radial shape larger than the junction area 
is patterned by EBL. Devices on this wafer are used to study 
the \emph{geometrically assisted cooling} effect.

A fifth wafer includes devices with a 120~nm Au cap of radial shape, patterned by EBL, 
its radius being $3~\mu$m wider than the junction radius (compare with devices 
\#6 and \#7 in Table~\ref{tab:01} and Figs.~\ref{fig:02}(e) and (f)). 
Like device \#5, these devices are also used to systematically study the effect of 
the \emph{geometrically assisted cooling} effect provided by the extended Au layer. 
In this device process the Au layer has a thickness of 80~nm. After the 
tunnel junction area definition and the subsequent CMP step, we observed 
during the measurement of the dielectric layer thickness that this layer was 
polished down to the top Nb electrode rather than having stopped at the 
80~nm thick Au layer. Therefore, for the subsequently patterned radial Au cap 
with thickness 120~nm, we assume equally a total thickness of 120~nm.

The fabrication of the dedicated SNS' devices (\#8 and \#9 in Table~\ref{tab:01}) 
for the discussion in Sec.~\ref{sec:05} uses a similar process as described before. 
This time the Au layer etch defines the junction area. We achieve a very small
device area by underexposing the poly(methyl methacrylate) (PMMA) resist which 
was used to define an aluminum nitride etch mask for the SNS' junction area definition.
The SNS' contact area could not be measured directly with our electron microscope 
due to its very small size and has to be determined over the normal state resistance 
together with a suitable theory. In this way we obtain SNS' device resistances of the 
order of $0.1-2~\Omega$. This allows to measure a sufficient large voltage drop 
across the device and, therefore, renders it possible to characterize the devices 
with our usual \emph{dc}-transport setup without the need of a SQUID read-out.
\subsection{Measurement setup}
\label{sec:03b}
\begin{figure}[t!]
\centering
\includegraphics[width=\columnwidth]{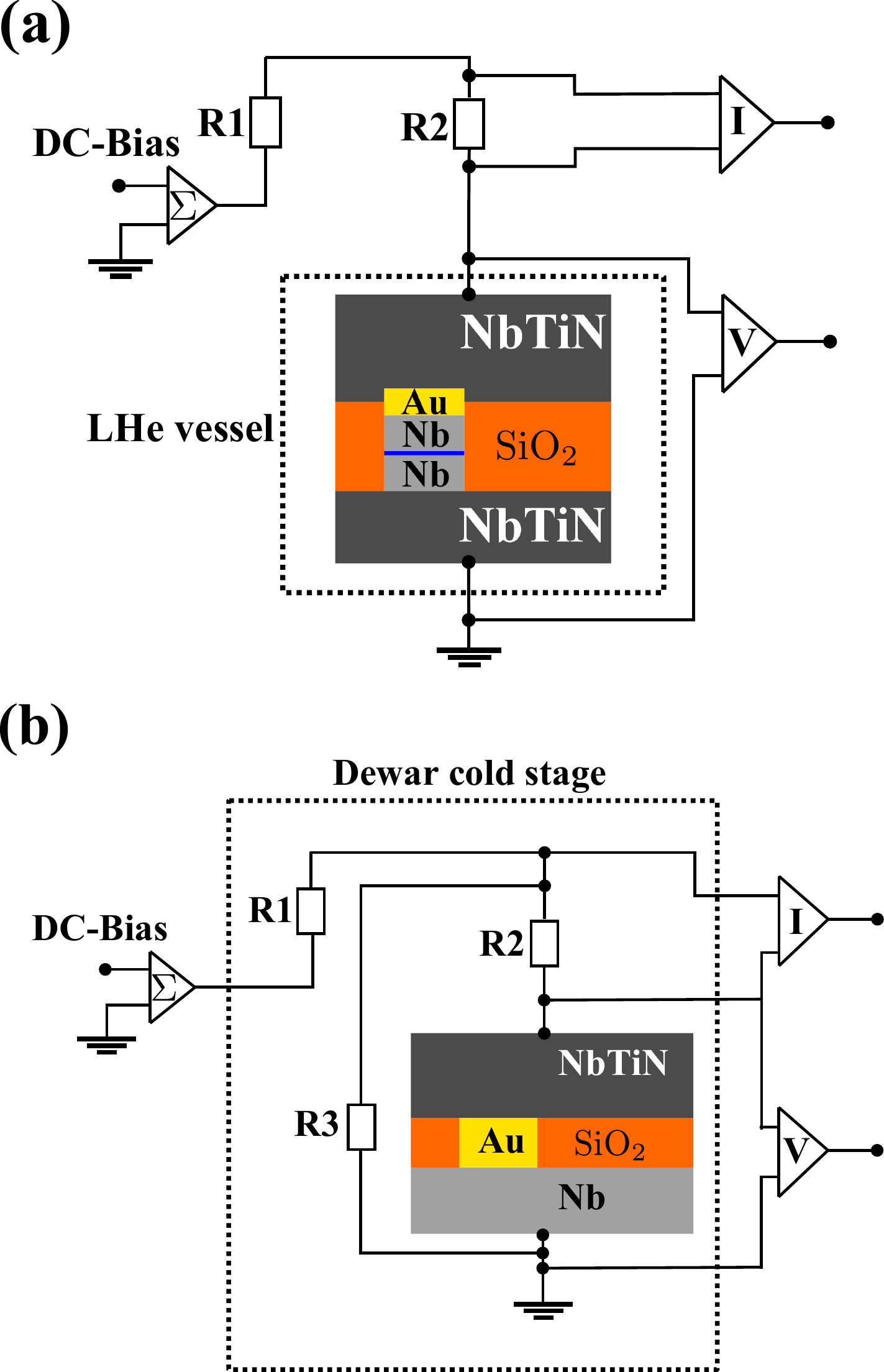}
\caption{(Color online) (a) Configuration for $\mathrm{S'SISS'}$ and $\mathrm{S'SISNS'}$ 
devices current-voltage measurements. The devices are directly immersed in LHe. (b) 
Configuration for $\mathrm{SNS'}$ current-voltage 
measurements with the devices mounted on the 4.2~K stage of a LHe dewar.}
\label{fig:03}
\end{figure}
The measurement setup for the device characterization is shown schematically in Fig.~\ref{fig:03}
for the S'SISS' and S'SISNS' devices \#1-\#7 (a) and the SNS' devices \#8 and \#9 (b). The S'SISS' 
and S'SISNS' devices in (a) are directly immersed in LHe in order to guarantee a device bath 
temperature of 4.2~K throughout the measurements of the different devices. On the other 
hand, the SNS' devices are characterized in a dedicated setup on the 4.2~K cold stage in 
a LHe dewar with bath temperatures of 4.24~K (device \#8) and 4.27~K (device \#9).

I,V characteristics of the S'SISS', S'SISNS' and the SNS' devices are measured with a 
resistively loaded current bias source. In order to measure a possible hysteresis in 
the SNS' I,V characteristic, we use a resistor ($R3$) connected in parallel to the SNS' junction.

The Josephson effect in the S'SISS' and S'SISNS' devices is suppressed by a magnetic field parallel to the 
junction barrier. The magnetic field strength is of the order of a 
few hundred Gauss corresponding to magnetic flux quanta in the junctions in the 
range of $\Phi_{0} - 2\Phi_{0}$, with $\Phi_{0} = h/2e$. 
During the measurements of the S'SISS' and S'SISNS' devices with applied magnetic field 
we did not observe a significant magnetic field induced energy broadening
of the quasiparticle density of states which would manifest in a weakened singularity 
in the I,V characteristic at the onset of the quasiparticle current branch at $V_{g} = 2\Delta/e$.

From this we conclude that the magnetic field did not significantly alter the 
superconductor's density of states. This is important for a realistic determination of the 
effective electron temperature $T_{e}$ which otherwise would result in lower $T_{e}$ 
values for too large magnetic fields.
\section{\label{sec:04}Experimental results and analysis}
On the basis of the conceptual background described in Sec.~\ref{sec:02}, we present in this 
section our experimental results obtained for the various devices \#1-\#7 shown in
Figs.~\ref{fig:02}(a)-(f), corresponding to Case I (normal metal Au layer is laterally confined to the 
same dimensions as the Nb junction) and Case II (normal metal Au layer is substantially 
wider than the Nb junction).

\begin{figure}[t!]
\centering
\includegraphics[width=0.8\columnwidth]{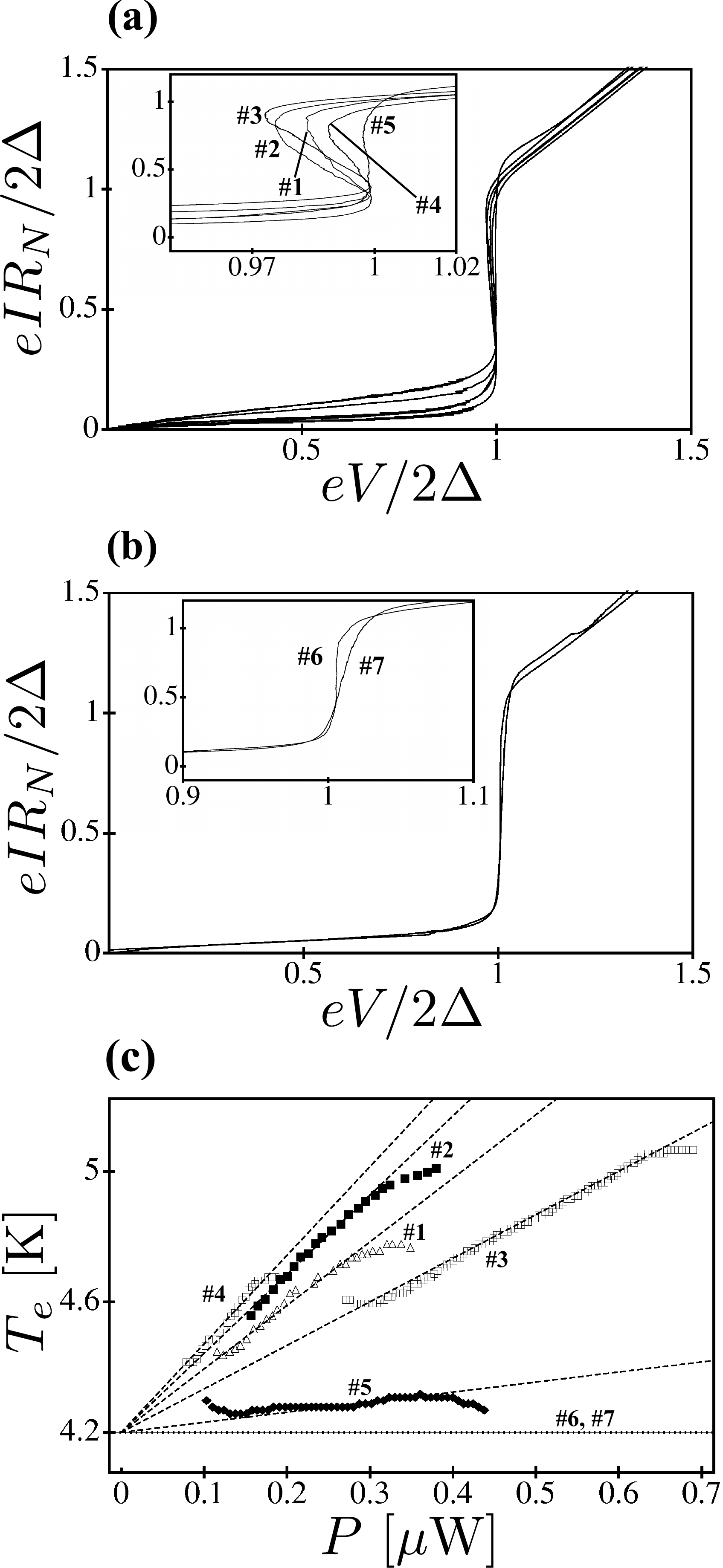}
\caption{(a) and (b) show the measured I,V characteristic of devices \#1-\#7
together with the measured effective electron temperature $T_{e}$ shown in (c). 
In (c), the $\triangle$ and $\blacksquare$-symbols show $T_{e}$ 
as a function of dissipated power in devices \#1 and \#2 whereas 
the $\Box$-symbol shows $T_{e}$ for devices
\#3 and \#4. The near thermal-bath electron temperature of device \#5 is 
indicated by the $\blacklozenge$-symbol. The horizontal 
dotted line designates the bath temperature of 4.2~K, which equally indicates the 
approximate electron temperatures of devices \#6 and \#7. Dashed lines are fits to the 
experimental data. Due to the different junction areas and current densities in the devices, 
the dissipated power spans different ranges. In (a) we observe an 
elevated electron temperature $T_{e}$ above the bath temperature 
$T_{ph}$, suggested by the back-bending feature (inset of (a)) of the 
quasiparticle current branch at $eV/2\Delta = 1$. 
This feature is weaker in (b).}
\label{fig:04}
\end{figure}

For the devices \#1 - \#4 which belong to Case I, we show the measured I,V curves in 
Fig.~\ref{fig:04}(a). The I,V characteristic of device \#5, belonging to Case II, is shown
in the same figure for comparison. Figure~\ref{fig:04}(b) shows the I,V characteristic of 
devices \#6 and \#7, also belonging to Case II. With the measured I,V curve as input parameter, we 
determine the effective electron temperature $T_{e}$ as a function of dissipated 
power $P = IV$ in the junction via Eq.~(\ref{eq:03}) over the temperature dependent gap 
energy $\Delta(T_{e})$. Results are shown in Fig.~\ref{fig:04}(c) for the various devices.

For the analysis of Case I, the heat-transfer coefficient $\alpha$ in Eq.~(\ref{eq:04}) 
can be extracted from a linear fit to the experimental curves
in Fig.~\ref{fig:04}(c) where $\alpha$ is equal to the inverse of the slope. 
On the other hand, $\alpha$ can be theoretically estimated and is determined 
principally by the heat-flow bottleneck in the device.

If the heat-flow bottleneck is between the electrons and the phonons, this means 
$\tau_{e-ph} > \tau_{es}$, the heat-transfer coefficient reads
$\alpha_{th} = V C_{e}/\tau_{e-ph}$, with $V$ being the 
volume of the junction and $C_{e}$ is the electronic heat capacity. On the other 
hand, if the heat-flow bottleneck is between the phonons and the bath, this is the case 
when $\tau_{e-ph} < \tau_{es}$, the heat-transfer coefficient reads 
$\alpha_{th} = V C_{ph}/\tau_{es}$. Here, $C_{ph}$ is the Debye phonon heat 
capacity, $C_{ph} = 234(T_{ph}/\theta_{D})^{3} n k_{B}$,\citep{ashcroft1975} with $\theta_{D}$
the Debye temperature and $n$ is the atomic density.

Since for Nb measurements of $\tau_{e-e}$ and $\tau_{e-ph}$ are rare and not avaliable for the 
film dimensions used in our devices, we resort to the results of 
\emph{Gershenzon et al.},\citep{gershenzon1990_01} $\tau_{e-e}\approx 0.1$~ns 
and $\tau_{e-ph} \approx 1$~ns at 4.2~K, which we assume to match 
best to our problem. Measurements of $\tau_{e-e}$ in Au revealed discrepancies of up to
four orders of magnitude to theoretical predictions.\citep{baselmans2001} 
Furthermore, experimental results from different groups showed significant differences in $\tau_{e-e}$. 
This suggests that for Au the method of sample fabrication and the material parameters have a 
large influence on the electron-electron interaction time in this particular material. 
Since for the analysis in this work it is only important to justify a short enough time 
$\tau_{e-e}$ (compare with Sec.~\ref{sec:02}), 
these discrepancies are not relevant for our analysis.
Thus, for the temperature regime of interest in this work and 
by considering the quality of our gold films (Table~\ref{tab:04}) we make use of the measurement 
results of \emph{Bergmann et al.}\citep{bergmann1990} on Au films in the dirty limit, obtaining
$\tau_{e-e} \approx 20$~ps and $\tau_{e-ph}\approx 40$~ps at 4.2~K. 

\begin{table}[t!]
\caption{\label{tab:02}Summary of theoretical and experimental values of the 
heat-transfer coefficient $\alpha$ for devices with increasing Au layer thickness, 
shown in Figs.~\ref{fig:02}(a), (b) and (c). The devices are directly immersed in LHe.
The electronic heat capacities for Nb and Au are evaluated 
for $T_{ph} = 4.2$~K via the relation $C_{e} = \gamma T$,\citep{kittel1957} whereas 
$C_{e}^{Nb}$ has to be corrected for the superconducting gap in the electronic 
quasiparticle spectrum.\citep{corak1956} The phonon-escape times are evaluated 
for a Nb-LHe and for a Au-LHe interface for various layer thicknesses.
$V_{Nb}$ and $V_{Au}$ are the volumes of the Nb electrodes and the 
additional Au layer on top of the junction. The two values in the "$\alpha_{th}$``-column are
obtained when using the relation 
$\alpha_{th} =  V_{Nb} (C_{e}/\tau_{e-ph}) + V_{Au} (C_{ph}/\tau_{es})$ 
(first value) or the relation 
$\alpha_{th} =  V_{Nb} (C_{e}/\tau_{e-ph}) + V_{Au} (C_{e}/\tau_{e-ph})$ 
(second value).}
\begin{ruledtabular}
\begin{tabular}{l l l l l}
\multicolumn{5}{l}{$C_{e}^{Nb} = 2000~\mathrm{\frac{J}{K m^{3}}}$; 
$C_{e}^{Au} = 300~\mathrm{\frac{J}{K m^{3}}}$}\\
\multicolumn{5}{l}{$\tau_{e-ph}^{Nb}= 1$~ns; $\tau_{e-ph}^{Au} = 40$~ps}\\ \\
\multicolumn{5}{l}{$C_{ph}^{Nb} = 630~\mathrm{\frac{J}{K m^{3}}}$; 
$C_{ph}^{Au} = 2870~\mathrm{\frac{J}{K m^{3}}}$}\\
\multicolumn{5}{l}{$\tau_{es}^{Nb}= 0.4$~ns (100~nm); $\tau_{es}^{Au} = 2.6$~ns (80~nm); $\tau_{es}^{Au} = 0.6$~ns (20~nm)}\\ \\
\hline
\#&$V_{Nb} \left[\mu\mathrm{m}^{3}\right]$&$V_{Au} \left[\mu\mathrm{m}^{3}\right]$
&$\alpha_{exp}~\left[\mathrm{WK^{-1}}\right]$&$\alpha_{th}~\left[\mathrm{WK^{-1}}\right]$\\
\hline
1&0.450&-&$5.1\cdot 10^{-7}$&\multicolumn{1}{c}{$9.0\cdot 10^{-7}$}\\
2&0.446&0.045&$4.1\cdot 10^{-7}$&$1.1\cdot 10^{-6}$; $1.2\cdot 10^{-6}$\\
3&0.504&0.2&$7.5\cdot 10^{-7}$&$1.2\cdot 10^{-6}$; $2.5\cdot 10^{-6}$\\
4&0.140&0.056&$3.7\cdot 10^{-7}$&$3.4\cdot 10^{-7}$; $7.0\cdot 10^{-7}$\\
\hline
\end{tabular}
\end{ruledtabular}
\end{table}
Because the S'SISS' and S'SISNS' devices are directly immersed in LHe, the escape times $\tau_{es}$ in 
Table~\ref{tab:02} are calculated using the measured Kapitza resistance 
between Nb and LHe\citep{amrit2000} and between Au and LHe.\citep{johnson1963}
From the original works, for Nb we use the relation 
$Y_{K} = (9.3\cdot T^{3.55} \cdot 10^{2})~\mathrm{W/m^{2}K}$ 
and for Au the relation $Y_{K} = (12\cdot T^{3} \cdot 10^{2})~\mathrm{W/m^{2}K}$ 
where $Y_{K}$ is the Kapitza conductance. 
Evaluating the relation $Y_{K} = (C_{ph}/\tau_{es})d$ at $T_{ph}=4.2$~K, 
with $d$ being the thickness of the particular device layer (Nb or Au), 
the phonon-escape time $\tau_{es}$ is obtained.

By comparing the phonon-escape time $\tau_{es}$ with the 
electron-phonon relaxation time $\tau_{e-ph}$, we conclude that the heat-flow bottleneck 
in the Nb is between the electrons and the phonons, whereas in the Au 
the heat-flow bottleneck is determined by the phonon-escape time to the LHe bath. 
Therefore, the theoretical heat-transfer coefficient $\alpha_{th}$ consists of two
different contributions, $\alpha_{th} =  V_{Nb} (C_{e}/\tau_{e-ph}) + V_{Au} (C_{ph}/\tau_{es})$, 
with $V_{Nb,Au}$ the volume of the two different layers in the junction. This value is shown
as the first number in the $\alpha_{th}$ column of Table~\ref{tab:02}. The second number 
in the same table indicates the heat-transfer coefficient when it is determined solely over 
the relation $\alpha_{th} = V (C_{e}/\tau_{e-ph})$ for both materials Nb and Au and is 
provided for comparison.

\begin{figure*}[t!]
\centering
\includegraphics[width=1\textwidth]{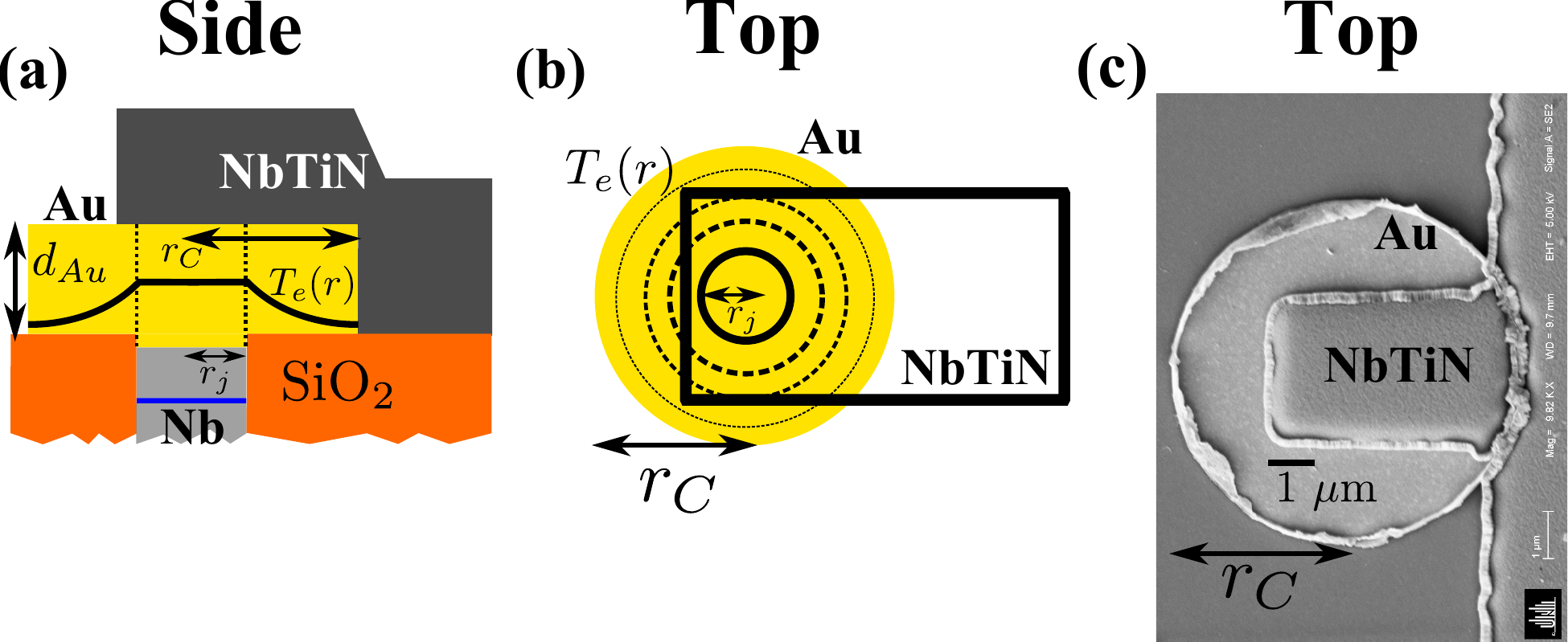}
\caption{(Color online) (a) side view of the upper part of a device like 
in Figs.~\ref{fig:02}(d)-(f). The thick black solid line indicates the 
decaying temperature profile $T_{e}(r)$ in the Au cap (compare with 
insets of Figs.~\ref{fig:06}(a) and (b)), illustrating the 
\emph{geometrically assisted cooling} effect. (b) sketches the top view 
of the device. The black rectangle is the NbTiN wiring layer. Dashed 
circles are surface lines of the radial temperature distribution 
$T_{e}(r)$, where the temperature gradient is indicated by the 
line thickness. The black solid line circle shows the contour of the SIS junction. 
(c) top view scanning electron microscope photograph 
(R.~Bruker, Institut f{\"u}r Physikalische Chemie, Universit{\"a}t 
zu K{\"o}ln) of one of our devices 
with an Au cap having radius $r_{C}>r_{j}$.}
\label{fig:05}
\end{figure*}
Considering first the results for $\alpha_{th}$ in Table~\ref{tab:02}, 
obtained for devices \#1 - \#3, a clear trend is observable.
Since all three devices have almost equal Nb junction volumes with values between 
$V_{Nb} = 0.450 - 0.504~\mu\mathrm{m}^{3}$, the effect of increasing Au layer thickness is 
to also increase the heat-transfer coefficient $\alpha_{th}$. This trend 
is consistent with the measurement of $\alpha_{exp}$ for devices \#2 and \#3.
The trend of $\alpha_{exp}$ for devices \#1 and \#2 is different from the theoretical 
prediction $\alpha_{th}$. Since for these two devices, the values $\alpha_{exp}$ are 
almost equal, we expect that the 20~nm Au layer of device \#2 
is thiner than expected or that the measurable cooling effect of the 
nonequilibrium quasiparticles is too small. 
Thus, no decrease in the electron temperature is observed.
For device \#4 the smallest heat-transfer coefficient is measured due to the smallest 
junction volume among all devices. In this device $\alpha_{th}$ and $\alpha_{exp}$ are in 
very good agreement, again assuming that the phonon-escape time in the Au is the slowest 
process. On the other hand, like for the devices \#1 - \#3, a larger deviation between 
$\alpha_{th}$ and $\alpha_{exp}$ is obtained by assuming electron-phonon scattering 
as the slowest process in the Au.

Taking account of the fact that we assume
constant parameters $C_{e}$, $\tau_{e-ph}$, $C_{ph}$ and $\tau_{es}$ for our interpretation, 
all values for $\alpha_{exp}$ and $\alpha_{th}$ are remarkably close together.
For example, in device \#1, doubling $\tau_{e-ph}$ leads already to
a match between the theoretical and the experimental value for $\alpha$. 
The best explanation for this difference between $\alpha_{exp}$ and $\alpha_{th}$ 
is the energy broadening of the quasiparticle density of states, expressed 
through the phenomenological parameter $\Gamma$, which we neglect in this analysis. 
This energy broadening leads to a smaller value of $T_{e}$ when directly 
extracted from the experimental data using Eq.~(\ref{eq:03}) and as 
a consequence of that to an underestimation of $\alpha_{exp}$ when 
fitting the $P$-$T_{e}$ data.
\begin{table}[t!]
\caption{\label{tab:03}Kapitza conductance $Y_{K}$, electron-phonon heat-transfer coefficient $Y_{e-ph}$, 
thermal conductivity $\kappa$ and thermal healing length $\eta$ for Au. 
Two values for $Y_{K}$ are specified. 
For the interface of Au and $\mathrm{SiO_{2}}$ measured at 4.2~K\citep{holt1966} 
and for the interface of Au and LHe, evaluated by the experimentally found 
relation 1.2$\cdot 10^{3} T^{3}$ at $T = 4.2$~K.\citep{johnson1963}
For the Au-LHe interface we use the lower Kapitza conductance in 
\emph{Johnson et al.}\citep{johnson1963} for a conservative estimate. 
$d_{Au}$ specifies the thickness of the Au cap on top of
the junction. $\eta_{1}$ includes the values $Y_{K}$, whereas $\eta_{2}$ is evaluated using $Y_{e-ph}$
for comparison. The radii $r_{C}$ of the Au caps of devices \#5-\#7 and the junction radii $r_{j}$ 
are indicated for comparison with $\eta$.}
\begin{ruledtabular}
\begin{tabular}{l l l l}
$Y_{K}~[\mathrm{W m^{-2}K^{-1}}]\times 10^4$:&\multicolumn{2}{l}{2.5~(Au-$\mathrm{SiO_{2}}$); 8.9~(Au-LHe)}\\
$Y_{e-ph}~[\mathrm{W m^{-2}K^{-1}}]\times 10^4$:&\multicolumn{2}{l}{45~($d_{Au}$=60~nm); 90~($d_{Au}$=120~nm)}\\
$\kappa~[\mathrm{W m^{-1}K^{-1}}]$:&\multicolumn{2}{l}{7}\\
\\
$\eta_{1}~[\mu\mathrm{m}]:$&\multicolumn{2}{l}{4.1~(Au-$\mathrm{SiO_{2}}$, $d_{Au}$=60~nm)}\\
&\multicolumn{2}{l}{5.8~(Au-$\mathrm{SiO_{2}}$, $d_{Au}$=120~nm)}\\
&\multicolumn{2}{l}{2.2~(Au-LHe), $d_{Au}$=60~nm)}\\
&\multicolumn{2}{l}{3.1~(Au-LHe), $d_{Au}$=120~nm)}\\
$\eta_{2}~[\mu\mathrm{m}]:$&\multicolumn{2}{l}{1}\\
\\
$r_{C}~[\mu\mathrm{m}]:$&\multicolumn{2}{l}{2.61~(\#5, $d_{Au}$=60~nm)}\\
&\multicolumn{2}{l}{4.71~(\#6, $d_{Au}$=120~nm)}\\
&\multicolumn{2}{l}{3.59~(\#7, $d_{Au}$=120~nm)}\\
$r_{j}~[\mu\mathrm{m}]:$&\multicolumn{2}{l}{0.61~(\#5)}\\
&\multicolumn{2}{l}{1.71~(\#6)}\\
&\multicolumn{2}{l}{0.59~(\#7)}\\
\end{tabular}
\end{ruledtabular}
\end{table}

For the analysis of Case II, we have to supplement Eq.~(\ref{eq:05}) 
with sufficient boundary conditions in order to describe the power 
input to the device and the balancing of heat via the aforementioned 
energy relaxation processes between the quasiparticles.

The following boundary condition describes the power dissipation in the SIS junction 
due to the bias current\citep{dieleman1996}
\begin{equation}
\label{eq:06}
-\kappa \frac{dT_{e}(r)}{dr} = \frac{IV}{2\sqrt{A\pi}d_{Au}},
~\quad\mathrm{for}~r\rightarrow r_{j}=\sqrt{\frac{A}{\pi}}~.
\end{equation}
Equation~(\ref{eq:06}) assumes a constant temperature in the Au layer within a cylinder of area 
$A$, equal to the SIS junction area, and 
height $d_{Au}$ equal to the thickness of the Au layer 
directly on top of the SIS junction. The cylinder range is indicated by the dotted lines in Fig.~\ref{fig:05}(a). 
This assumption is justified since it turns out that $\eta$, the characteristic length of Eq.~(\ref{eq:05}) 
(thermal healing length), is larger than any other relevant layer thickness in the device, hence suggesting
a constant temperature in the junction. Thus, this boundary condition accounts for the 
power outflow per unit area from the surface of the cylinder 
(the SIS junction areas $A$ are summarized in Table~\ref{tab:01}).

The thermal conductivity $\kappa$ at $T = 4.2$~K of the Au layer 
is determined via the Wiedemann-Franz law, $\kappa = LT/\rho$, 
using the measured resistivity $\rho$ of Au at 4.2~K, given in Table~\ref{tab:04}, 
and the Lorenz number for Au in Ref.~25\nocite{ashcroft1975}, taken to be $L = 2.32~\mathrm{W\Omega/K^2}$.
Then the thermal healing length can be defined as $\eta = [(\kappa d)/Y]^{1/2}$.\citep{skocpol1974} 
This length has to be compared to the radius $r_{C}$ of the Au caps in our 
devices \#5-\#7 (compare with Figs.~\ref{fig:02}(d)-(f)). All values are given in Table~\ref{tab:03}.

Since $\eta\approx r_{C}$ in our devices, the second boundary condition for Eq.~(\ref{eq:05}) reads
\begin{equation}
\label{eq:07}
\frac{dT_{e}(r)}{dr} = 0,~\quad\mathrm{for}~r\rightarrow r_{C},
\end{equation}
rather than $T_{e} \rightarrow T_{ph}$ for $r\rightarrow r_{C}$, which is suitable only for large radii 
$r\gg\eta$ and would allow to determine analytical solutions.

\begin{figure*}[t!]
\centering
\includegraphics[width=1\textwidth]{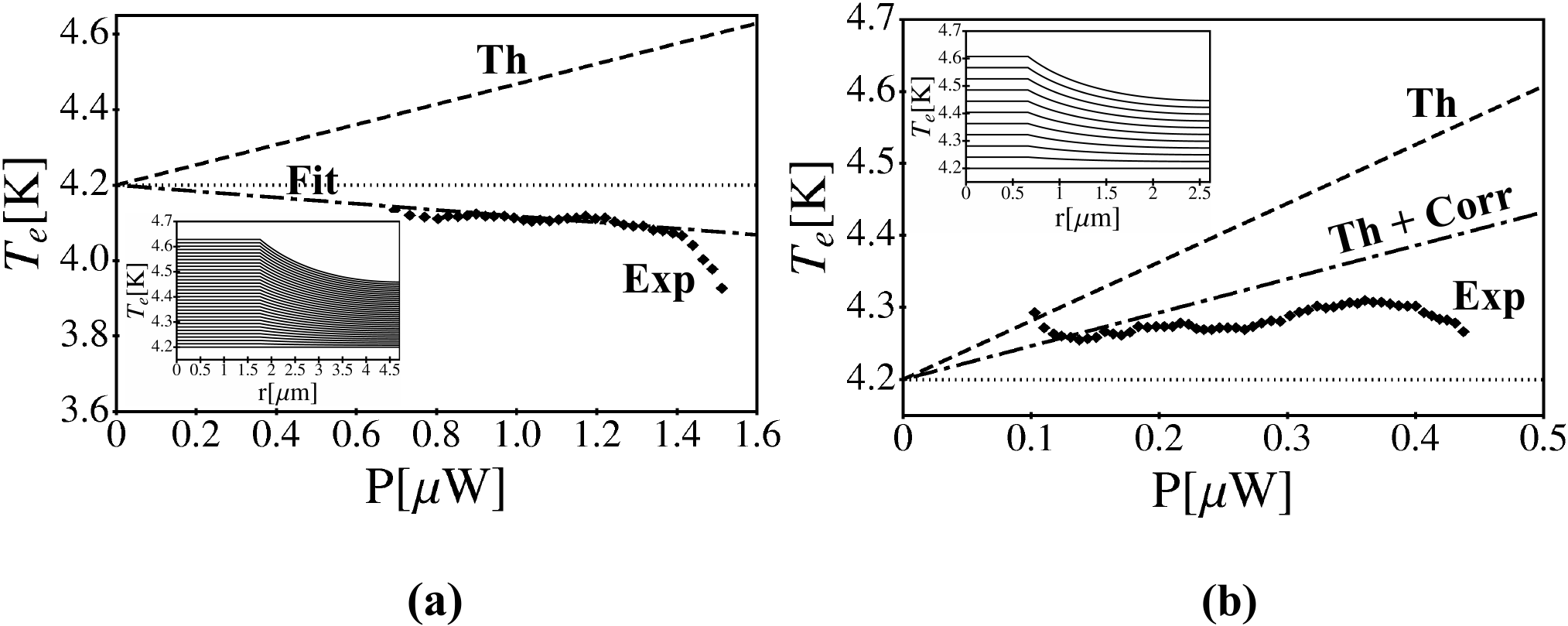}
\caption{Solutions of Eq.~(\ref{eq:05}) (curves labeled with "Th") together with 
experimental results (curves labeled with "Exp"). (a) shows the result for 
device \#6 whereas (b) shows the result for device \#5. 
In (b) two theoretical predictions are given. The dashed line ("Th``) shows the 
solution within the Au cap volume directly on top of the SIS junction in a 
range $r\in [0,r_{j}]$ (compare with the values in Table~\ref{tab:03}).
This solution does not include the correction due to the energy broadening 
of the quasiparticle density of states which we finally 
extract from a fit in (a) (curve labeled with "Fit``). 
The dashed-and-dotted curve in (b) shows 
the modified trend ("Th+Corr") with this correction included. The insets 
present the spatial dependence of $T_{e}$ over the Au cap as a function 
of the radial distance $r$ for various levels of dissipated power in the 
junction in steps of $0.05~\mu$W. Here, the position $r = 0~\mu$m indicates 
the middle of the Au cap. The straight line at 4.2~K in the inset figures 
is the solution for $P = 0~\mu$W for which $T_{e} = T_{ph}$. The horizontal dotted lines
in (a) and (b) indicate the phonon temperature $T_{ph}$.}
\label{fig:06}
\end{figure*}
We arrive at the following assumptions for the analysis presented below.
First, we assume that the Nb-Au-NbTiN system (wiring part of the device) has 
clean interfaces, i.e.~no or only very thin oxide layers between the different metals. 
Therefore, these metals share one common phonon system. Furthermore, the thermal 
healing length is determined only by the phonon-escape rate at the Au-LHe interface 
when the devices are directly immersed in LHe or by the phonon-escape rate at the 
Au-$\mathrm{SiO_{2}}$ interface when we consider devices 
which are cooled in a vacuum environment. With these assumptions we solve 
the heat balance Eq.~(\ref{eq:05}) and present our results for devices \#6 and \#5 in 
Figs.~\ref{fig:06}(a) and (b). For the calculations we use the Kapitza conductance 
at the Au-LHe interface since these devices were directly immersed in LHe
during our measurements.

We start first with the discussion of the obtained results in Fig.~\ref{fig:06}(a). 
The experimental results were obtained by evaluation of Eq.~(\ref{eq:03}) with the I,V values of 
device \#6 as input parameters. The dashed theoretical line shows the prediction of the 
electron temperature rise with increasing dissipated power in the Au cap volume 
directly on top of the SIS junction (region within the dotted lines in Fig.~\ref{fig:05}(a)). 
This result is obtained by the solution of Eq.~(\ref{eq:05}) within the range $r\in [0,r_{j}]$ and 
disregards a correction of $T_{e}$ due to the energy broadening of the quasiparticle density of states.
On the other hand, the slope of the experimentally determined $P$-$T_{e}$ curve is slightly negative 
because of the slightly positive slope of the quasiparticle current branch in the I,V characteristic 
of device \#6 (compare with Fig.~\ref{fig:04}(b)) due to the energy broadening of the 
quasiparticle density of states. Hence, we obtain an electron temperature
smaller than the bath temperature. The slope of this $P$-$T_{e}$ curve is solely used to 
extract the energy broadening correction of the quasiparticle density of states 
to the effective electron temperature $T_{e}$ and does not represent the real 
device temperature of device \#6. In device \#6, due to the largest area among all devices in 
this paper, the lowest magnetic field is necessary in order to suppress the \emph{dc} Josephson effect. 
We expect, therefore, that in this device the local density of states of the superconducting electrodes 
are least modified by the magnetic field. The fit to the data in this device is shown 
by the dashed-and-dotted line (labeled with "Fit``). As a result of the fit 
we find that the slope of the theoretical prediction 
(dashed line) can be mapped onto the experimental slope by adding a term 
$-0.35~\mathrm{K}/\mathrm{\mu W}$.

When adding the correction term $-0.35~\mathrm{K}/\mathrm{\mu W}$ 
to the theoretical prediction of the $P$-$T_{e}$ trend of device \#5, 
the curve labeled with "Th+Corr" in Fig.~\ref{fig:06}(b) is 
obtained. We find that we can describe the data sufficiently well following this approach. 
The small discrepancy between the curve labeled with "Th+Corr" and the experimental curve 
labeled with "Exp" can be best explained by a small modification of the local density 
of states of the superconducting Nb electrodes by the magnetic field. 
Because of the significant smaller cross section of device \#5 
compared to device \#6 (compare with the junction radius $r_{j}$ in Table~\ref{tab:03}), 
a larger magnetic field has to be applied to device \#5 to induce the same magnetic flux 
as in device \#6. Therefore, by adding an additional conservative 
correction term with value $-0.15~\mathrm{K}/\mathrm{\mu W}$ to the "Th+Corr" curve 
(not shown in the figure), accounting for the influence of the magnetic field on the 
local density of states, we can describe the experimentally obtained slope sufficiently 
well over the full range of the dissipated power $P$.

For device \#7, we did not perform a similar analysis like described before. 
During the measurements with this device we applied the strongest magnetic field 
among all other measurements, necessary due to the small
junction area of this device (cf.~Table~\ref{tab:03}). Hence the 
experimental $P$-$T_{e}$ slope which one would obtain from the corresponding I,V curve 
should not be considered to be a representative measurement.
However, we show the I,V characteristic of this device in Fig.~\ref{fig:04}(b) for 
comparison to the I,V characteristic of device \#6. We observe that the slope at the
onset of the quasiparticle branch of the I,V characteristic of device \#6 is almost vertical whereas for 
device \#7 we measure an I,V characteristic with a positive slope.

Although the experimental result for device \#6 (and in principle also for device \#7) 
suggests that the effective electron temperature has reached the phonon temperature, 
we have to stress that the effective electron temperature in the device 
will follow the theoretical prediction (dashed curve) in Fig.~\ref{fig:06}(a) as this trend 
does not contain the energy broadened quasiparticle density of states which effectively 
decreases the experimental result for $T_{e}$.
This means that the effective electron temperature in device \#6 is still slightly 
above the phonon temperature. Comparing the results for device \#6 and \#5 in 
Figs.~\ref{fig:06}(a) and (b) we find that at the same dissipated power of say
$0.2~\mu\mathrm{W}$, the effective electron temperature in device \#6 is about 
0.1~K lower than in device \#5. This once more shows the \emph{geometrically 
assisted cooling effect} of the Au layer on the electron temperature $T_{e}$ in the Nb electrodes.

When calculating $T_{e}$ in device \#5 with an Au cap having thickness $d = 120$~nm and
$r_{C} = r_{j} + 3\mu\mathrm{m}$, therefore, choosing the same geometry as for the Au cap in devices \#6 and \#7,
we find that at $0.4~\mu$W dissipated power, $T_{e} \approx 4.24$~K 
(including the correction for the quasiparticle density of states energy broadening).
Thus, this electron temperature is approximately 60~mK lower than the measured value for the actual device
geometry. At some value of $r_{C}$, a further increase of the Au cap radius has no influence 
anymore on decreasing $T_{e}$ and $d_{Au}$ has to be thicker. Evidently, dependent on 
how much power is dissipated in the device, there are many design possibilities 
in order to achieve an electron temperature in the device equal/close to the phonon temperature.

Since the thermal healing length $\eta$ is independent of the dissipated power in the device and of 
the Au cap radius $r_{C}$, it can only be used for a rough estimate of the Au cap size. However, 
as a necessary condition for the \emph{geometrically assisted cooling} we find that obviously $r_{C} > \eta$. 
Nevertheless, even when this condition is fulfilled, in principle one has to solve Eq.~(\ref{eq:05}) 
with the specific device geometry as input parameter in order to determine 
the electron temperature in the device.
\section{\label{sec:05}Possible adverse effects due to inserted normal metal layer}
Before we present our conclusions we would like to make a few remarks which have to be
considered when implementing our device concept in a practical mixer design for a heterodyne 
receiver experiment. Here, it is important to estimate possible signal losses caused by the 
normal metal Au layer in the superconducting circuit.

First, from \emph{dc}-characterization measurements of the SNS' devices \#8 and \#9, not 
further discussed in detail in this paper, we determine a critical current of 
$I_{c} \approx 200~\mu$A at $T = 4.24$~K for device \#8 and a critical current 
of $I_{c} \approx 5$~mA at $T = 4.27$~K for the larger area device \#9, cf.~also Table~\ref{tab:01}. 
The normal state resistances of devices \#8 and \#9 are $R_{N} = 1.76~\Omega$ and $R_{N} = 0.082~\Omega$. 
A tunnel current through the S'SISNS' device larger than the critical current of the SNS' part 
causes that a series resistance appears which is 
equal to the normal state resistance of the SNS' part. However, the typical contact areas of the SNS' layer 
structure on top of the tunnel barrier are much larger than the areas of the test devices \#8 and \#9 which we 
fabricated extra-small in order to obtain a significant voltage-drop across the device, simplifying the measurements. 
Therefore, in our S'SISNS' devices, it is to be expected that the critical current of the SNS' 
part exceeds the tunnel current through the insulating barrier by far and, thus, for a \emph{dc} current 
the normal metal Au layer does not represent a series resistance to the overall S'SISNS' normal state resistance.

Second, in case of an \emph{rf} current induced by a microwave signal in a heterodyne experiment, the Au layer 
causes \emph{rf} signal loss. However, compared to the size of other circuit parts of typical SIS mixer layouts, the
micron-sized Au layer is comparatively small and should, therefore, not entail a significant amount of 
signal loss in the mixer. Nevertheless, in the design of high-Q tuning circuits for THz frequencies, 
crucial for low-noise mixer operation and for efficient coupling the signal radiation to the mixing element 
over a broad frequency bandwidth, the proximity-effect modified local density of states in the normal 
metal Au layer due to the two superconductors Nb and NbTiN, changing the complex conductivity 
of the normal metal material, has to be considered by a 
generalized \emph{Mattis and Bardeen} theory.\citep{nam1967,mattis1958}
\section{\label{sec:06}Conclusions}
To conclude, in this paper we have reported several device geometries combining the high-quality
Nb-trilayer technology with a full superconducting NbTiN embedding circuit and conducted 
\emph{dc}-characterization measurements of the devices. We measured the effective electron temperature $T_{e}$ 
as a function of the applied bias current in terms of the electron-temperature dependent 
superconducting gap energy $\Delta(T_{e})$. We find that a normal metal layer, in our experiments we used Au, 
sandwiched between the Nb junction and between one of the NbTiN leads, decreases the electron temperature
compared to the control device in which no normal metal layer is used. For a normal metal layer which is 
laterally confined to the same area like the Nb electrode (Case I), we measure an increase in the cooling
of the nonequilibrium quasiparticle system with increasing thickness $d_{Au}$ of the normal metal layer.
We identify two different heat-flow bottlenecks in these devices, namely between the electrons and 
the phonons in the Nb material, whereas in the Au the phonon escape process from the material to the 
bath is the slowest energy relaxation mechanism. The resulting theoretical heat-transfer coefficient
$\alpha_{th}$ describes the experimentally determined value $\alpha_{exp}$ sufficiently well. 
Case II covers devices which have normal metal Au layers substantially wider than the 
Nb electrodes. Here we observe enhanced cooling of the nonequilibrium quasiparticles compared 
to Case I which we relate to \emph{geometrically assisted cooling}, i.e.~nonequilibrium 
quasiparticles relax their energy in the wide Au layer by outdiffusion, electron-phonon 
scattering and phonon escape to the thermal bath. The heat-flow can be explained by a 
simple thermal model which takes the Au layer geometry and the aforementioned energy 
relaxation processes into account, provided that one considers also the energy broadening 
of the quasiparticle density of states. We have highlighted the prospects of using our 
device technology in a future heterodyne experiment where one can expect a substantial 
improvement in the mixer sensitivity due to the low \emph{rf}-loss property of the NbTiN
material and the high quality of the Nb-trilayer junction technology, compared to devices 
which implement at least one normal conducting layer in the embedding 
circuit.\citep{jackson2001,jackson2005}
\begin{acknowledgments}
This work is carried out within the Collaborative 
Research Council 956, sub-project D3, funded by 
the Deutsche Forschungsgemeinschaft (DFG) and  
by BMBF, Verbundforschung Astronomie under 
contract no.~05A08PK2. We thank the AETHER program of RadioNet~3 EU 
FP 7 for financial support. M.~P.~Westig 
thanks the Bonn-Cologne Graduate School of 
Physics and Astronomy (BCGS) for financial support 
of a half-year stay abroad at the Kavli Institute of 
Nanoscience in Delft, The Netherlands, and the CosmoNanoscience 
group in Delft for great hospitality during this time. The help of Dr.~M.~Justen 
during modification of the bias electronics is gratefully acknowledged.
M.~Schultz helped in dealing with the technical issues during sample measurements. 
The devices were fabricated in the KOSMA 
microfabrication laboratory and the measurements were 
conducted at the I.~Physikalisches Institut, 
Universit{\"a}t zu K{\"o}ln. We thank R.~Bruker from the Institut f{\"u}r 
Physikalische Chemie of the Universit{\"a}t zu K{\"o}ln for the 
SEM picture in Figure~\ref{fig:05}(c).
\end{acknowledgments}
\appendix
\section*{\label{app:01}Material parameters}
\begin{table*}[ht]
\caption{\label{tab:04}Material parameters for the layers of devices 
\#1-\#9 in Table~\ref{tab:01}. For each device and material, the first, 
second and third number indicates the resistivity $\rho$, measured 
at 20~K, the critical temperature $T_{c}$ of the superconducting material 
and the superconducting pair breaking energy $2\Delta$. Only for the Au layer, 
the resistivity was measured at 4.2~K for a 200~nm thick film on a separate wafer.}
\begin{ruledtabular}
\begin{tabular}{l l l l l}
\#&NbTiN(b)&Nb&Au&NbTiN(t)\\
\hline
\multirow{3}{*}{1}&133~$\mu\Omega$cm&5.9~$\mu\Omega$cm&-&140~$\mu\Omega$cm\\
&14.45~K&9.12~K&-&14.37~K\\
&4.38~meV&2.69~meV&-&4.36~meV\\
\\
\multirow{3}{*}{2}&-&5.9~$\mu\Omega$cm&1.3~$\mu\Omega$cm&125~$\mu\Omega$cm\\
&14.53~K&9.07~K&-&14.58~K\\
&4.41~meV&2.67~meV&-&4.42~meV\\
\\
\multirow{3}{*}{3,4}&133~$\mu\Omega$cm&5.9~$\mu\Omega$cm&1.3~$\mu\Omega$cm&140~$\mu\Omega$cm\\
&14.55~K&9~K (\#3); 9.33~K (\#4)&-&14.55~K\\
&4.41~meV&2.65~meV (\#3); 2.76~meV (\#4)&-&4.41~meV\\
\\
\multirow{3}{*}{5}&128~$\mu\Omega$cm&5.9~$\mu\Omega$cm&1.3~$\mu\Omega$cm&128~$\mu\Omega$cm\\
&14.86~K&9.04~K&-&14.56~K\\
&4.51~meV&2.66~meV&-&4.41~meV\\
\\
\multirow{3}{*}{6,7}&172~$\mu\Omega$cm&5.9~$\mu\Omega$cm&1.3~$\mu\Omega$cm&147~$\mu\Omega$cm\\
&14.52~K&9.15~K (\#6); 9.12~K (\#7)&-&14.04~K\\
&4.41~meV&2.7~meV (\#6); 2.69~meV (\#7)&-&4.26~meV\\
\\
\multirow{3}{*}{8}&-&5.9~$\mu\Omega$cm&1.3~$\mu\Omega$cm&182~$\mu\Omega$cm\\
&-&8.72~K&-&14.23~K\\
&-&2.53~meV&-&4.31~meV\\
\\
\multirow{3}{*}{9}&-&7~$\mu\Omega$cm&1.3~$\mu\Omega$cm&145~$\mu\Omega$cm\\
&-&8.63~K&-&14.48~K\\
&-&2.49~meV&-&4.39~meV
\end{tabular}
\end{ruledtabular}
\end{table*}
Table~\ref{tab:04} summarizes all material parameters of the devices studied 
in this paper. For the NbTiN layers, we determine the superconducting pair breaking 
energy from the measurement of the film critical temperature $T_{c}$ by 
numerical inversion of the integral equation (3.27) of 
\textit{Bardeen et al.} \citep{bardeen1957} where we assume 
$\Delta(0) = 1.764\cdot k_{B}T_{c}$. The integral is carried out up to the 
Debye temperatures $\theta_{D} = 275$~K for Nb\citep{kittel1957} 
and $\theta_{D} = 310$~K for NbTiN. The latter value of $\theta_{D}$ is determined 
via a fit of the Bloch-Gr\"uneisen theory to the measured NbTiN resistivity 
as a function of temperature. For the Nb junction material in devices \#1-\#7, 
we measure the superconducting pair breaking energy via the gap voltage $V_{g}$ of the SIS 
I,V characteristic, assuming $V_{g} = 2\Delta/e$.
Resistivities are determined in a four-terminal measurement 
configuration. For device \#2, the resistivity of the bottom wiring layer
could not be measured due to a defective four-point film structure.
The Nb and NbTiN superconducting pair breaking energies of the SNS' devices
\#8 and \#9 are determined over the critical temperature as described before.

The Au layer resistivity was not measured individually for each device. 
For all of our devices, we assume the measured resistivity value at 
4.2~K for a 200~nm thick film, sputter deposited on a separate wafer.
\providecommand{\noopsort}[1]{}\providecommand{\singleletter}[1]{#1}%
\end{document}